\RequirePackage{fix-cm}
\documentclass[smallextended]{svjour3}
\smartqed

\usepackage{amsmath,amssymb,mathrsfs}
\usepackage{graphicx}
\usepackage{graphics}
\usepackage{epsfig}
\usepackage{color}
\usepackage{nicefrac}

\newcommand{\beq}{\begin{equation}}
\newcommand{\eeq}{\end{equation}}
\newcommand{\beqa}{\begin{eqnarray}}
\newcommand{\eeqa}{\end{eqnarray}}

\renewcommand\Re{\operatorname{Re}}
\renewcommand\Im{\operatorname{Im}}

\DeclareMathOperator{\diag}{diag}

\begin{document}

\title{Universality in cuprates: a gauge approach}
\subtitle{}

\author{P.A.~Marchetti$^{1,2}$ \and G.~Bighin$^{1,2}$}

\institute{ \email{marchetti@pd.infn.it, bighin@pd.infn.it} \\
	        $^1$ Dipartimento di Fisica e Astronomia ``Galileo Galilei'', 
	        Universit\`a di Padova, Via Marzolo 8, 35131 Padova, Italy \\
	        $^2$  Istituto Nazionale di Fisica Nucleare, Sezione di Padova, 
	       Via Marzolo 8, 35131 Padova, Italy }

\date{Received: date / Accepted: date}

\maketitle

\begin{abstract}

In high-$T_c$ cuprates many quantities exhibit a non-Fermi liquid 
universality hinting at a very peculiar structure of the underlying 
pairing mechanism for superconductivity: in this work we focus on the universality for the in-plane 
resistivity and the superfluid density.

We outline the previously developed  spin-charge gauge approach to 
superconductivity in hole-doped cuprates: we decompose the hole of the $t-t'-J$ model for the $\mathrm{Cu}\mathrm{O}_2$ planes as the 
product of a spinful, chargeless gapped spinon and a spinless, 
charged holon with Fermi surface. Each one of these particle excitations 
is bound to a statistical gauge flux, allowing one to optimize their 
statistics.

We show that this model allows for a natural interpretation of the 
universality: within this approach, under suitable conditions, the spinonic and
holonic contributions to a response function sum up according to the 
Ioffe-Larkin rule. We argue that, if the spinonic contribution 
dominates, then one should
expect strongly non-Fermi-liquid-like universality, due to the 
insensitivity of spinons to Fermi surface details. The in-plane 
resistivity and
superfluid density are indeed dominated by spinons in the underdoped 
region. We theoretically derive these quantities, discussing their 
universal behaviours  and comparing them with experimental data.

\end{abstract}

\section{Introduction}

The interpretation of the low-energy physics of high-$T_c$ cuprates has not yet reached a common consensus. In this general mystery one of the puzzling features is the appearance of universal behaviour of some physical quantities when suitably normalized, with strongly non-Fermi-liquid-like character. These quantities exhibit, within some region of the phase diagram bounded by crossovers or phase transitions, independence from both doping concentration and the specific kind of material involved. Two typical quantities showing this feature in underdoped hole-doped cuprates are the in-plane resistivity $\rho_\|$ in the normal state and the superfluid density $\rho^{(s)}$ in the superconducting state. Let us make this statement more precise. 
We denote by $T^*$ the (lower) pseudogap temperature identified for example by the inflection point in the in-plane resistivity and by $T_{\text{MIC}}$ the temperature corresponding to the metal-insulator crossover as defined by the minimum of in-plane resistivity in underdoped samples. Then the normalized resistivity defined by
 \begin{equation}
\label{res}
\rho_{\|n} (T) =\frac{\rho_\| (T)-\rho_\| (T_{\text{MIC}})}{\rho_\| (T^*)-\rho_\| (T_{\text{MIC}})} 
\end{equation}
exhibits an universal behaviour when expressed as a function of the normalized temperature $T/T^*$, as first discussed in Ref. \cite{wuyts} and also observed in Refs. \cite{trappeniers,xiang}.
Analogously, if $T_c$ denotes the temperature for the onset of superconductivity, then the superfluid density, $\rho^{(s)}$ normalized as
\begin{equation}
\label{sup}
\rho^{(s)}_n (T/T_c) =\frac{\rho^{(s)} (T/T_c)}{\rho^{(s)} (T=0)} 
\end{equation}
also shows an universal behaviour, as noticed in YBCO samples \cite{hardy} and discussed for a wide variety of materials in Ref. \cite{mb}. A systematic analysis of universality in other physical quantities can be found in Ref. \cite{xiang}.

In this paper we show that the universality discussed above can be naturally explained and explicitly reproduced within a gauge approach to the low-energy physics of hole-doped cuprates developed in Refs. \cite{msy,mfsy}. This approach is based on modelling the $\mathrm{Cu}\mathrm{O}_2$ planes of hole-doped cuprates in terms of a $t-t'-J$ model and it strongly relies on a composite nature of the holes, corresponding to the Zhang-Rice singlets in physical materials. The hole is viewed as a resonance obtained binding together through gauge fluctuations an excitation with Fermi surface carrying the charge of the hole but spinless, the holon, and a neutral gapped spin $\nicefrac{1}{2}$ excitation, the spinon. As shown below this composite nature of the hole is at the origin of the non-Fermi-liquid universal behaviour discussed in this paper, which can appear if the spinon dominates the physical quantity one is considering, because in this situation no Fermi surface details are present in the response. 

The plan of the paper is the following: in Section 2 we outline the key ideas of the spin-charge gauge formalism and in Section 3 its application to superconductivity. In Section 4 we discuss some fundamental conditions for the emergence of the non-Fermi-liquid universality and in Sections 5 and 6 we discuss the application to in-plane resistivity and superfluid density, respectively.

\section{Spin-charge gauge approach to hole-doped cuprates}
 In this formalism the hole field is at first seen as a product of a bosonic spinon field, $z$  and a fermionic holon field, $h$, which, being spinless, implements exactly the Gutzwiller constraint of no-double occupation of the $t-t'-J$ model by Pauli principle.

This decomposition introduces an unphysical degree of freedom
due to a local $U(1)$ gauge invariance, because one can
multiply at each site the spinon and the holon by arbitrary
opposite phase factors leaving the physical hole field
unchanged. This invariance is made manifest with the
introduction of a slave-particle gauge field $A$, which, in
turn, produces an attraction between $z$ and $h$. At this level the long wavelength continuum limit of the model is described by a Fermi liquid of holons  and an O(3) non-linear model for spinons, coupled by the gauge field.

At the second stage, however, we add a $\nicefrac{1}{2}$ charge flux $\Phi_h$ to the holon and a $\nicefrac{1}{2}$ spin flux $\Phi_s$ to the spinon, still retaining the fermionic statistics
of the hole. This is materialized in the Lagrangian formalism coupling the holon to a charge- and the spinon to a spin-Chern-Simons gauge field, $B$ and $V$, gauging the global $U(1)$-charge and $SU(2)$-spin invariances of the model, respectively.
One can rigorously prove \cite{fm} that the coupling to the gauge fields with coefficients -2 for $B$ and 1 for $V$ does not change the physical content of the model. Let us outline the key idea of the proof for the partition function. We expand the partition function of
the gauged model in the first-quantized formalism in terms of
the world lines of holes. It turns out that the effect of the coupling to the charge- (spin-) gauge fields is only to give a
factor $e^{\pm i \pi/2}( e^{\mp i \pi/2})$ for any single exchange of the hole
world lines, so the two effects cancel each other exactly.

The coupling to the Chern-Simons gauge fields turns both holons and spinons into semions, i.e. particle excitations obeying the braid statistics, which can be characterized by the phase factor $e^{\pm i \pi/2}$ of the many-body wave-function when two semions are exchanged.
The reason for this change of statistics is two-fold: first, in the 1-dimensional model this statistics is crucial to get the correct critical exponents, second, in two dimensions it is crucial to get the correct Fermi surface for the hole. In fact, experimentally for optimally doped materials one finds a Fermi surface in agreement with band calculations. However, this is clearly troublesome for fermionic spinless holons  because being spinless their mean-field Fermi volume is expected to be doubled w.r.t. to the hole case \cite{lee2}. 

On the other hand, a semion in this model, as conjectured in Ref. \cite{msy2} and recently proved in Ref. \cite{fmsy}, obeys a Haldane exclusion statistics \cite{haldane} with parameter $\nicefrac{1}{2}$, implying that for low temperature one can accommodate at most two (spinless) semions in the same momentum state and the semion distribution function at $T \approx 0$ is twice the fermion distribution function. Hence a gas of such  spinless semions of finite density has a Fermi surface coinciding with that of spin $\nicefrac{1}{2}$ fermions of the same density. In the approximate treatment developed in Ref. \cite{msy} after taking into account this exclusion effect any semionic character of the holons and spinons is neglected. A more careful treatment of the semionic features is presently under investigation \cite{fmsy}, but here we follow the approximate treatment. Even within this approximation scheme, universality and a good agreement with experimental data emerge for the quantities we consider here and elsewhere \cite{mosy,mdosy,mfsy}.

We consider a mean-field (MF) approach where we neglect
the holon fluctuations in $\Phi_h$ and the spinon fluctuations
in $\Phi_s$. 
This leads to a much simpler form
of the two statistical fluxes. The charge one is actually static
and it provides a $\pi$-flux phase factor per plaquette. As a consequence of Hofstadter mechanism this flux
converts the low-energy modes
of the spinless holons $h$ into Dirac fermions with dispersion
defined in the magnetic Brillouin zone and a small
Fermi surface  $\epsilon_F \sim t \delta$, 
characterizing what we dub the ``pseudogap phase'' (PG) of
the model. Increasing doping or temperature, one reaches the
crossover line $T^*$ quoted in the Introduction.
 Crossing this line, we enter
in the ``strange metal phase'' (SM) in which the effect of the
charge flux is screened by the background spinon configuration
in MF approximation and we recover a "large" tight-binding Fermi surface for the holons with
$\epsilon_F \sim t (1 +\delta)$. Below $T^*$ the $t'$ term is essentially irrelevant due to the appearance of a small Fermi surface, on the other hand it is needed above $T^*$ to reproduce the correct tight-binding Fermi surface. Since in the present paper we only work in pseudo-gap regime, the strongly material-dependent $t'$ term is not relevant for the results we discuss here. For the SU(2) spin flux in MF approximation only the
third component survives and we have:
\begin{equation}
\label{flux}
\Phi_s(j) =(\sigma_3/2) \sum_l i \arg(j - l)h^*_l h_l (-1)^{|l|},
\end{equation} 
 where $\sigma$ denotes the Pauli matrices, $j, l$ are lattice sites
 and $|l| = l_x +l_y$. The gradient of $\Phi_s$ can be viewed
as the potential of $U(1)$ spin vortices centered at holon positions
with $(-1)^{|l|}$ chirality. Hence the empty sites of the 
model, mimicking the Zhang-Rice singlets and corresponding to the holon locations, are the cores of spin vortices, a quantum distortion of the antiferromagnetic  spin
background, with opposite chirality in the two N\`eel sublattices.
These vortices appear in the U(1) subgroup of the $SU(2)$
spin group complementary to the coset labeling the directions
of the spin \cite{pam}. Fluctuations of such directions describe the
spin-waves, viewed as composites of spinons generated by
gauge attraction between spinon and antispinon of the $O(3)$ continuum model. Therefore,
the spin-vortices have a purely quantum origin, behaving somewhat
analogously to the flux in the Aharonov-Bohm effect.
The additional interaction term between spinons and
spin-vortices in the continuum limit is of the form
\beq
\label{spvo}
J(1 - 2\delta)( \nabla \Phi^s(x))^2 z^*(x)z(x)
\eeq
and it is the source of both short-range AF and charge
pairing. In fact, from a quenched treatment of spin-vortices, we
derive the MF expectation value $\langle ( \nabla \Phi^s(x))^2 \rangle = m^2_s
˜\approx 0.5 \delta | \log \delta|$, which opens a mass gap for the spinons, consistent
with AF correlation length for small dopings, as extracted from
the neutron experiments \cite{keimer}. Thus, propagating in the gas
of slowly moving spin-vortices, the spinons, originally
gapless in the absence of spin vortices, acquire a finite
gap, leading to a short range AF order. 

\section{Superconductivity in spin-charge gauge approach}
By averaging the spinons instead of
the spin flux in Eq. (\ref{spvo}), we obtain an effective interaction:
\beq
\label{coul}
J(1 - 2 \delta) \langle z^*z \rangle \sum_{i,j} (-1)^{|i|+|j|} \Delta^{-1}
 (i - j) h^*_ih_i h^*_jh_j,
\eeq
where $ \Delta$ is the two-dimensional lattice Laplacian.
This pairing is the driving force for superconductivity. The corresponding order parameter, $\Delta^h$, turns out to be $d$-wave and the system is not far from the BCS-BEC
crossover, but still on the BCS side \cite{mb}. Since the pairing
originates from spin-vortices it is independent of nesting
features of the Fermi surface, used in most spin-wave
approaches.
From Eq. (\ref{coul}), we see that
the interaction mediated by spin-vortices on holons is of
2D Coulomb type. From the known behaviour of planar
Coulomb systems, we derive that below a crossover temperature
$T_{ph} ˜\sim J(1 - 2\delta) \langle z^*z \rangle$  a finite density of incoherent holon pairs appears.
This charge pairing does not yet lead to hole-pairing,
since the spins are still unpaired. The hole-pairing is
achieved by the gauge attraction between holon and spinon.
The holon-pairs as source of attraction
lead to the formation of short-range spin-singlet
(RVB) spinon pairs. Hence, at an intermediate
crossover temperature lower than $T_{ph}$ and denoted
by $T_{ps}$, a finite density of incoherent spinon RVB pairs
appears; combined with the holon pairs, they give rise to a
gas of incoherent preformed hole pairs.

The lowering of
free energy allowing the formation of spinon pairs is a consequence
of the origin of the spinon gap from screening
due to unpaired vortices, so that the short-range vortex-antivortex
pairs essentially do not contribute to it. Therefore, with a
mechanism clearly not BCS-like, the spinon
gap lowers proportionally to the density of spinon pairs,
implying a lowering of the kinetic energy of spinons.
The doping density
enters in the above mechanism through two factors: the
density of hole pairs and the strength of the attraction
behaving like $˜\sim J(1 - 2\delta)\langle z^*z \rangle$, as seen from Eq. (\ref{coul}). These two effects
act in an opposite way increasing doping, thus yielding a
"dome" shape to $T_{ph}(\delta)$, starting from a non-zero doping
concentration.
Superconductivity (SC) occurs by condensation of hole pairs
at a temperature $T_c$ lower than $T_{ps}$ inheriting the dome
structure.

\begin{figure}
  \includegraphics[scale=0.65]{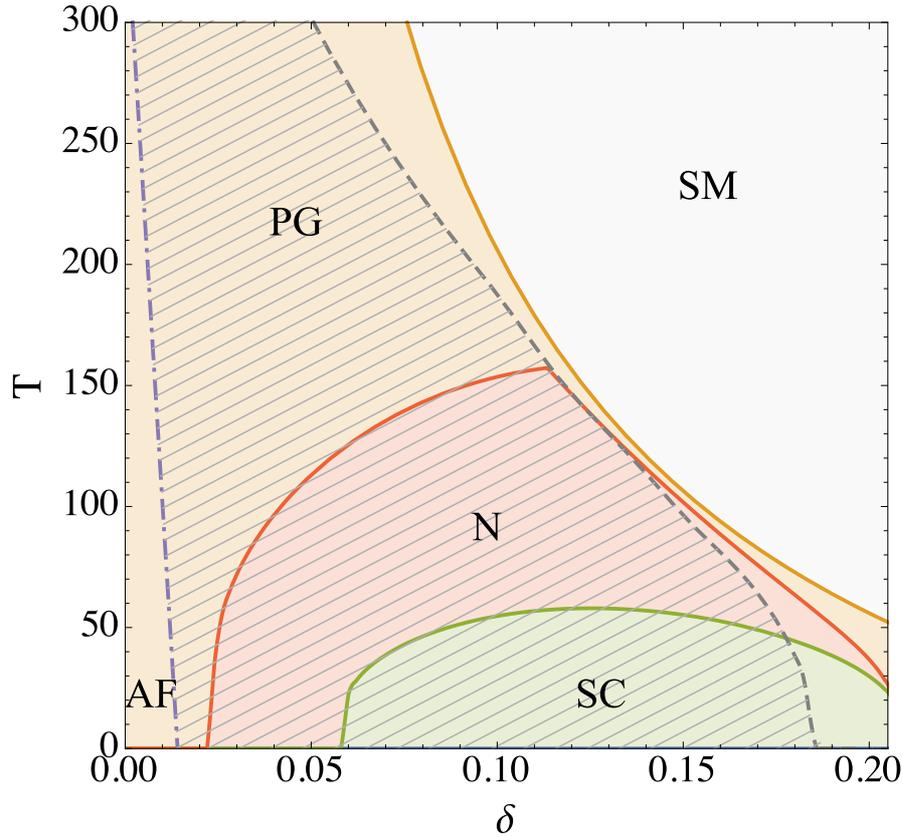}
\caption{Theoretically derived phase diagram: the holon pairing temperature $T_{ph}$ (yellow line) is determined from the gap equation for holons in the BCS approximation. Similarly the spinon pairing temperature $T_{ps}$ (red line) is determined from the spinon gap equation and encloses the N region in which the system supports a Nernst signal. The crossover PG-SM, denoted by the dashed line, is determined above $T_{ps}$ from the inflection point of resistivity and below $T_{ps}$ from matching the contour lines of the values of the spinon order parameter derived from the gap equation for spinons. $T_c$ (green line) is determined from the transition temperature of the XY model of spinons. The N\`eel temperature (dot-dashed line), delimiting the region characterized by anti-ferromagnetic (AF) order, is qualitatively obtained from experiments and has not yet been derived theoretically in our approach. The shaded region is the region analyzed in this paper.}
\label{fig:0}       
\end{figure}

We dub ``Nernst'' (N) the region in the phase diagram between $T_{ps}$ and $T_c$,
since the gap of preformed pairs allows for the formation of magnetic vortices
supporting a Nernst signal even in the absence of superconductivity \cite{ong}.
The phase diagram as derived from the formalism just outlined is shown in Fig. \ref{fig:0}.

\section{Origin of universality}

From the outline of the gauge approach presented above it is clear that whereas the holon retains detailed information about the Fermi surface of the hole, the spinon does not. Its propagator still depends on the doping concentration, via $m_s$ and indirectly via the holon susceptibility, appearing through the coupling to the gauge field, but it does not depend on details of the Fermi surface.
At a general level it is then natural to conjecture that if the behaviour of a physical quantity is controlled by the spinons, then there is a chance to find universality with strongly non-Fermi liquid character.
To find such spinon-dependent behaviour we need three conditions: 1) the quantity we compute depends essentially on spinons and holons by themselves, not on the hole as a whole, 2) the spinon contribution is the dominating one, 3) the characteristic temperature is set by spinons.

For quantities depending on the low energy-momentum electromagnetic polarization bubbles condition 1) holds if the Ioffe-Larkin composition rule \cite{ioffe} is valid. In turn this rule holds if the spinon-gauge and the holon-gauge systems have an independent scaling limit and the corresponding gauge effective actions are Gaussians, i.e. RPA approximation is the leading term in the scaling limit. Actually both systems are expected to have this feature, the holons being Fermi-liquid and the spinons being massive.
Let us sketch the derivation of the Ioffe-Larkin rule, which is only a consequence of gauge-invariance under the assumptions made above.
We couple the electromagnetic field $A_{em}$ to holons by minimal coupling, then denoting by $\Pi^h_{\mu\nu} ( \Pi^s_{\mu\nu})$ the polarization bubbles of the holon-gauge (spinon-gauge) system in the scaling limit and the effective action $S(A_{em})$ for $A_{em}$ in that limit can be obtained by:
\begin{eqnarray}
\label{SAem}
e^{-S(A_{em})}= \int {\cal D}A e^{- \frac{1}{2} (A^\mu+A^\mu_{em})\Pi^h_{\mu\nu}(A^\nu+A^\nu_{em})- \frac{1}{2} A^\mu\Pi^s_{\mu\nu}A^\nu}=\nonumber\\
e^{- \frac{1}{2} A^\mu_{em}(\Pi^h(\Pi^h+\Pi^s)^{-1}\Pi^s)_{\mu\nu}A^\nu_{em}}.
\end{eqnarray} 
Hence the electromagnetic polarization bubble is given by
\beq
\label{IL}
\Pi_{em}= \Pi^h(\Pi^h+\Pi^s)^{-1}\Pi^s
\eeq
and this is the Ioffe-Larkin composition rule.
If $\Pi^h \gg \Pi^s$, then $\Pi_{em} \approx \Pi^s$ and the condition 2) of spinon dominance is satisfied.
Notice that if there are no parity breaking terms in the polarization bubbles and we work in the Coulomb gauge for $A$, then Eq. (\ref{IL}) holds separately for the transverse,  $\Pi_\bot$ and the temporal $\Pi_0$ polarization bubbles.

\section{Universality in resistivity}

Resistivity is the most peculiar feature of a superconductor: a sudden resistivity drop when reaching the critical temperature is, in fact, the defining feature of superconductivity.

Resistivity in cuprates in the normal state above the critical temperature, along with many other experimental features, is anisotropic and has a completely different behaviour when measured along the $\mathrm{Cu}\mathrm{O}_2$ planes (in-plane resistivity) or measured in a perpendicular direction (out-of-plane resistivity). In this work we focus on the in-plane resistivity, an analysis of the off-plane resistivity within the present formalism can be found in Ref. \cite{mdosy}.

The in-plane resistivity $\rho_\|$ in cuprates exhibits quite peculiar behaviour: in the normal state, i.e. above the critical temperature, it evolves from a metallic-like behaviour $\rho_\| \sim T^2$ in the strongly-overdoped regime to a linear behaviour $\rho_\| \sim T$ at the optimal doping. In the underdoped regime in-plane resistivity increases less than linearly and it either drops to zero at the critical temperature, or diverges for $T=0$ if the doping is so low that there is no superconductivity down to $T=0$. For sufficiently underdoped samples one can identify a minimum in resistivity at $T_{\text{MIC}}$ marking a metal-insulator crossover. At higher temperatures, an inflection point in the resistivity as a function of temperature is often used in defining the lower pseudogap temperature $T^*$. On the other hand, the upper pseudogap temperature is identified by the deviation of $\rho_\|$ from the linear behaviour.

We now show that the spin-charge gauge approach introduced in the previous sections can reproduce the experimental behaviour of the in-plane resistivity in the normal state for the pseudo-gap regime, within good approximation, revealing also its universality. 

As the hole is decomposed into the product of a spinful, uncharged spinon and a spinless, charged holon one expects two different contributions to resistivity. 
Since the resistivity is obtained from the polarization bubble through the Kubo formula \cite{kubo}:
\beq
\label{Kubo}
(\rho_i)^{-1}  = 2 \Re \int_0^{\infty} \mathrm{d} x^0 x^0 \Pi^i_\bot (x^0, \mathbf{q} = 0)
\eeq
with $i=h,s,em$ and $\rho_{em}=\rho_\|$, the Ioffe-Larkin rule applies with the result:
\beq
\rho_\| = \rho_s + \rho_h \; ,
\label{eq:ilresist}
\eeq
where $\rho_s$ ($\rho_h$) is the contribution from the spinon (holon) subsystem, respectively. 

Let us briefly sketch the derivation of the polarization function and of the spinon in-plane resistivity within the present formalism \cite{mdosy}. The spinon in-plane dynamics in the low-energy continuum limit is described by a non-linear $\sigma$ model \cite{mfsy}
\beq
\label{o3}
S_s = \frac{1}{g} \int \mathrm{d}^3 x \left[ v_s^{-2} |(\partial_0 - \mathrm{i} A_0) z_\alpha|^2 + |(\partial_i - \mathrm{i} A_i) z_\alpha |^2 + m_s^2 z^*_\alpha z_\alpha \right],
\eeq
where the massive  $O(3)$ model has been written in terms of the gauge field, with $g \sim J^{-1}$ and $v_s \sim J a$, $a$ being the lattice spacing. The integration in Eq. (\ref{o3}) can be extended to the full three-dimensional Euclidean space-time $\mathbb{R}^3$, treating the imaginary time component and the space components on the same footing, since in the range considered $v_s /T \sim J/T \gg 1$, as also noted in Ref. \cite{mb}. However, we retain the temperature dependence of the coefficients in the action. The spinon propagator in  external gauge field can be written in the Schwinger representation as
\beq
G_\alpha (x,y | A) = \langle z(x) z^* (y) \rangle (A) = \mathrm{i} g v_s \int_0^\infty \mathrm{d} s \ e^{-\mathrm{i} s (\Delta_A + m_s^2)} (x,y) \; ,
\eeq
where $\Delta_A$ is the $(2+1)$-dimensional Laplacian and the zeroth (time) coordinate has been rescaled by a factor $v_s$.
After having calculated the spinon Green function one can derive the polarization function as
\beq
\Pi_s (x-y) = \langle D_{A(x)} G(x,y|A) D^\dagger_{A(y)} G(y,x|A) \rangle
\eeq
and $D_{A(x)} $ is the covariant derivative. Integrating over the spatial coordinates one formally obtains $\Pi_s(x^0, \mathbf{q}=0)$.  An approximate analytical approach was proposed in Ref. \cite{mdosy}. One takes into acount  the effect of the gauge field in the eikonal approximation. Its typical momentum scale is given by the Reizer momentum $Q_0(T) \approx (\kappa T/ \chi)^{1/3}$, where $\kappa \sim \delta$ is the Landau damping and $\chi \sim 1/\delta$ the diamagnetic susceptibility of holons. In the scaling limit the main effect of gauge fluctuations is to renormalize the mass of the spinon introducing a damping factor: 
\beq
m_s \longrightarrow \sqrt{m_s^2-\frac{i c T}{\chi}} \; ,
\eeq
where $c \approx 3$, and to introduce a scale $\sim Q_0^{-1}$ of average distance between spinon and holon. The space and time integrals involved in the computation of the Kubo formula in Eq. (\ref{Kubo}) are approximately computed by scale renormalization and principal value evaluation, and are dominated by a saddle point at the typical scale quoted above in the range
\begin{equation}
   m_s^2 \gtrsim \frac{T}{\chi}\gtrsim m_s Q_0.
\end{equation}
In physical units, this gives a range of temperatures between a
few tens and a few hundreds of Kelvin and we identify this region in the $\delta-T$ phase diagram with the experimental region between the lower pseudogap crossover and the spin-glass phase; this is the region we dub PG.
The output of the  above computation is
\beq
\rho_s^{-1} \sim \Im(\sqrt{\kappa}(m_s^2 - i c\frac{T}{\chi})^{1/4}).
\eeq
One can rewrite the result \cite{mdosy}, making explicit the doping dependence,  as:
\beq
\label{rs}
\rho_s \approx  \sqrt{\frac{m_s}{\delta}} \frac{\left[ 1+ \left( \frac{\xi}{\lambda_T} \right)^4 \right]^{\frac{1}{8}}}{\sin \left[ \frac{1}{4} \arctan^2 \left( \frac{\xi}{\lambda_T} \right) \right]}
\eeq
with $\xi=m_s^{-1} \sim |\delta \ln \delta|^{-\frac{1}{2}}$, $\lambda_T \sim (\chi / T c)^{\frac{1}{2}}\sim (\delta T )^{-\frac{1}{2}}$.
For low $T$, $\rho_s \sim {1\over T}$, thus
exhibiting an insulating behaviour, for $T \gtrsim \chi m_{s}^2$
one finds $\rho_s \sim T^{1/4}$, thus showing a metallic
behaviour. From Eq. (\ref{rs}) a metal-insulator crossover is thus recovered
decreasing the temperature.  This crossover is determined by the interplay between
the AF correlation length $\xi$ and the thermal de Broglie wave
length $\lambda_T$. In the  limit $\lambda_T \geq \xi$ we find a ``peculiar''
localization due to short-range AF order, producing an
insulating behaviour (due to the gauge interaction $\rho_s
\neq e^{{(\frac{\Delta}{T})}^\alpha}$, a behaviuor found for a
``standard'' localization). When $\lambda_T \lesssim \xi$ this
localization is not felt and a metallic behaviour is
observed. 
The holon contribution can be found as in Ref. \cite{mdosy} by adapting the calculation in Ref. \cite{lee1} and introducing the contribution from impurities through the Matthiessen rule
\beq
\rho_h \sim \delta \left[ (\epsilon_F \tau)^{-1} + \left( \frac{T}{\epsilon_F} \right)^{\frac{4}{3}} \right]
\eeq
where $\tau$ is the characteristic scattering time due to impurities, allowing one to calculate the total in-plane resistivity through Eq. (\ref{eq:ilresist}). For small $\delta$, $T/t$ we have $\rho_s \gg \rho_h$, so the spinon contribution dominates the physical resistivity, thus satisfying condition 2) of the previous Section.
At last we normalize the resistivity as follows
\beq
\rho_{\| n}(T/T^*) = \frac{\rho_\|(T/T^*) - \rho_\|(T_{\text{MIC}}/T^*)}{\rho_\|(T^*/T^*) - \rho_\|(T_{\text{MIC}}/T^*)} \; .
\label{eq:rhon}
\eeq
The characteristic temperature $T^*$ is essentially determined by spinons, since it corresponds to the inflection point in $\rho_s$, thus satisfying the condition 3) for universality.
In fact, neglecting the holon contribution the normalized resistivity as defined in Eq. (\ref{eq:rhon}) would be a universal function of the normalized temperature $T/T^*$. 

\begin{figure}
  \includegraphics[scale=0.75]{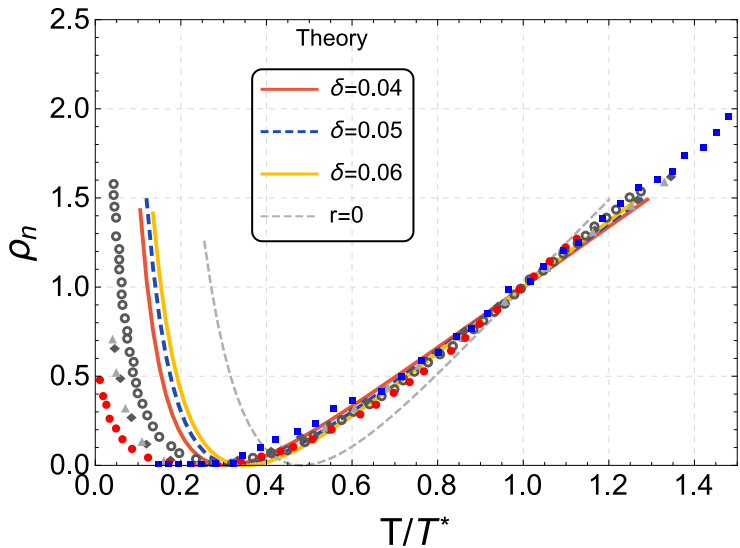}
\caption{The in-plane resistivity as theoretically calculated for $\delta=0.04$, $\delta=0.05$, $\delta=0.06$, shows near-universality when normalized as analyzed in the main text; our results are compared with experimental data for YBCO and LSCO from Refs. \cite{ando,takagi} and for $\mathrm{Bi}_2 \mathrm{Sr}_{1.6} \mathrm{La}_{0.4} \mathrm{Cu} \mathrm{O}_y$ from Ref. \cite{kostantinovic}. Gray open circles denote the doping-independent behaviour of $\mathrm{La}_{2-x} \mathrm{Sr}_x \mathrm{Cu} \mathrm{O}_4$ (LSCO) for $x=0.04$ and $x=0.05$ in Ref. \cite{takagi}; gray triangles and diamonds denote experimental data from $\mathrm{La}_{2-x} \mathrm{Sr}_x \mathrm{Cu} \mathrm{O}_4$ (LSCO), for $x=0.03$ and $x=0.04$, respectively, taken from Ref. \cite{ando}; red filled circles denote $\mathrm{Y} \mathrm{Ba}_2 \mathrm{Cu}_3 \mathrm{O}_y$ (YBCO) experimental data for $y=6.35$, taken from Ref. \cite{ando}; finally blue squares denote experimental data for $\mathrm{Bi}_2 \mathrm{Sr}_{1.6} \mathrm{La}_{0.4} \mathrm{Cu} \mathrm{O}_y$, with $p=0.06$, $p$ is the in-plane carrier concentration, from Ref. \cite{kostantinovic}. The $r=0$ curve corresponds to the universal, pure spinon contribution \cite{mdosy}. The discrepancy at low $T$ might be due to the missing account of holon and spinon pair formation in the above treatment.}
\label{fig:1}       
\end{figure}

The previous computation leaves as free parameter a relative coefficient $r$ between the holon and the spinon contribution, independent of both $T$ and $\delta$. The relative weight of the spinon and holon contributions to resistivity is not fixed due to the use of a scale renormalization in the continuum limit in the computation of the spinon contribution, hence the need for the free parameter $r$ fixing the relative weights. Adding the holon contribution with an optimized $r$, while adding a slight degree of non-universality, allows for a quantitatively correct fit of experimental data, as shown in Fig. \ref{fig:1} over a wide range of temperatures and dopings. In fact, although the holon contribution is essentially negligible at low $T$ it modifies the position of the inflection point $T^*$, improving the agreement with experiments of $\rho_{\| n}(T/T^*)$. Notice that $r$ is the only free parameter in the theoretical curves of Fig. \ref{fig:1}. To show that the qualitative structure of the curve is also independent of $r$, we also plot there the curve with $r=0$ corresponding to the pure spinon contribution, without free parameters \cite{mdosy}.

\section{Universality in superfluid density}

The superfluid density is a fundamental quantity in the description of a superconducting system: a naive figure would interpret it within a two-fluid model as the density of charge carriers taking part in the resistance-less current flow, while a more precise definition relates it to the coefficient of the term governing fluctuations of the phase $\phi$ of the order parameter in an effective action for superconductivity \cite{prokofev}:
\begin{equation}
S_\text{eff} = \frac{\rho_s}{2} \int \mathrm{d} V \left( \nabla \phi \right)^2 + \left( \text{other terms} \right) \; .
\label{eq:cuprates:effact0}
\end{equation}
From this definition and using the minimal coupling to the electromagnetic field $A_{em}$, the superfluid density is readily related to the London penetration depth
\beq
\lambda = \sqrt{\frac{m}{\mu_0 \rho_s e^2}} \; ,
\eeq
where $e$ and $m$ being the charge and the mass, respectively, of each charge carrier, $\mu_0$ being the vacuum permeability, implying that the superfluid density can can be indirectly obtained by measuring the exponential decay of an external magnetic field inside the superconductor. Hence it corresponds to the zero-frequency and zero-momentum term in $\Pi_{em \bot}$. 
As opposed to the resistivity which is non-zero only in the normal state, superfluid density is non-zero only in the superconducting phase, i.e. below the critical temperature. In particular, as also analyzed in Ref. \cite{mb}, the superfluid density has a quite peculiar behaviour: the critical exponent at $T_c$ is $\nicefrac{2}{3}$, putting it in the 3D XY universality class, while a linear $T$-dependence at low temperature seems reminiscent of a more standard BCS-like $d$-wave description; even though these feature seem at odds, they can be reproduced naturally within the present spin-charge gauge approach.

As seen for the resistivity, even for superfluid density one can separately calculate a spinon and a holon contribution, and again since $\rho^{(s)}$ is derived from  $\Pi_{em \bot}$ the  Ioffe-Larkin composition rule applies, thus realizing condition 1) for universality:
\beq
(\rho^{(s)})^{-1} = (\rho^{(s)}_s)^{-1} + (\rho^{(s)}_h)^{-1} \; ,
\label{eq:ilrhos}
\eeq
as analyzed in Ref. \cite{mb}. Again for moderate underdoping the spinon contribution is dominating, thus realizing condition 2).

The additional term to the Lagrangian of Eq. (\ref{o3}) arising from spinon pairing is of the form:
\beq
\label{spo}
\sum_{i=1,2}\Delta^s_{i}(x) \epsilon^{\alpha\beta}
z_{\alpha}\partial_{i} z_{\beta}(x) - \frac{|\Delta^s_0|^2}{J |\Delta^h_0|^2} \; ,
\eeq
where $\Delta^s_{i}$ is the spinon order parameter in the continuum limit, $\Delta^s_0$ its amplitude, assumed constant and $\Delta^h_0$ the amplitude of the holon order parameter, also assumed constant.
Below $T_{ps}$ the dynamics of the superconducting transition is described by an effective Lagrangian density, obtained by integrating out holons the spinons. Its leading term due to spinons is a three-dimensional anisotropic gauged XY model given by:
\beq
\mathscr{L}_{\text{eff}} = \frac{1}{6 \pi M} \{ [ \partial_\mu A_\nu - \partial_\nu A_\mu ]^2 + | \Delta^s_0 |^2 [ 2 ( A_0 + \partial_0 \frac{\phi}{2} )^2 + ( \mathbf{A} + \nabla \frac{\phi}{2} )^2 ] \} \; .
\eeq
where $\phi$ is the condensate phase and $M \approx m_s-| \Delta^s_0 |^2/m_s$. 
At the superconducting transition, the gauge field $A$ is
gapped by the Anderson-Higgs mechanism and the gauge
fluctuations are suppressed. 
Hence the gauge contribution turns out to be subleading \cite{mb} and, as far as the superfluid density is concerned, one can simply consider the three-dimensional XY model described by the following Lagrangian density
\beq
\mathscr{L}_{\text{XY}} = \frac{\left| \Delta^s_0 \right|^2}{6 \pi M} \eta^{\mu \nu} \partial_\mu \frac{\phi}{2} \partial_\nu \frac{\phi}{2}
\label{eq:lxy}
\eeq
with $\eta^{\mu \nu}=\diag(2,1,1)$, $\phi$ being the hole-pair condensate phase. It is important to note that the imaginary time component now plays the same role as the two spatial components, leading to a three-dimensional model which is not the result of inter-layer coupling, as sometimes advocated to interpret the experimental data. The condition allowing one to treat the imaginary-time component on the same footing as the spatial components has been introduced after Eq. (\ref{o3}) and is always satisfied in the temperature range we consider.

The effective inverse temperature
\beq
\Theta^{-1}(T) = \frac{\left| \Delta^s_0(T) \right|^2}{3 \pi M}
\label{eq:cuprates:defbigtheta}
\eeq
plays the role of the physical inverse temperature $\beta$ in the XY model. The behaviour of the model is not altered from a qualitative point of view, because $\Theta^{-1}(T)$ is a monotonically decreasing function of the temperature as $\beta$: the transition between the low-temperature and high-temperature phases of the three-dimensional XY model is now determined by $\left| \Delta^s_0 (T)\right|^2$. Hence again the spinons are determining the characteristic temperature, thus realizing condition 3) for universality. As common in case of condensates $\Delta^s_0 (T)/\Delta^s_0 (0)$ turns out to be a universal function of $T/T_{ps}$, $T_{ps}$ being the characteristic temperature of formation of spinon pairs. 
The spinon contribution to superfluid density at zero temperature is then
\beq
\rho^{(s)}_s(0) \approx  \left[\frac{\mathrm{d} \Theta}{\mathrm{d} T}(0) \right]^{-1}
\eeq
and the full temperature profile is recovered as
\beq
\rho^{(s)}_s(T) =\rho^{(s)}_s(0) \rho_{XY}(\Theta(T)/\Theta(T_c)) \; .
\eeq
Here $\rho_{XY}$ is the spin stiffness of the anisotropic 3D XY model in Eq. (\ref{eq:lxy}). Since $\Theta(T)$ is essentially linear in $T$, $\Theta(T)/\Theta(T_c) \approx T/T_c$ and $\rho_s^s(T) /\rho_s^s(0)$ is approximately the universal function
$\rho_{XY}(T/T_c)$ derived from the 3D classical XY model. Therefore the normalized spinon contribution to the superfluid density is near-universal, except for doping values extremely close to the edge of the superconducting dome, where linearity in $T$ of $\Theta$ breaks down.

\begin{figure}
\includegraphics[scale=0.75]{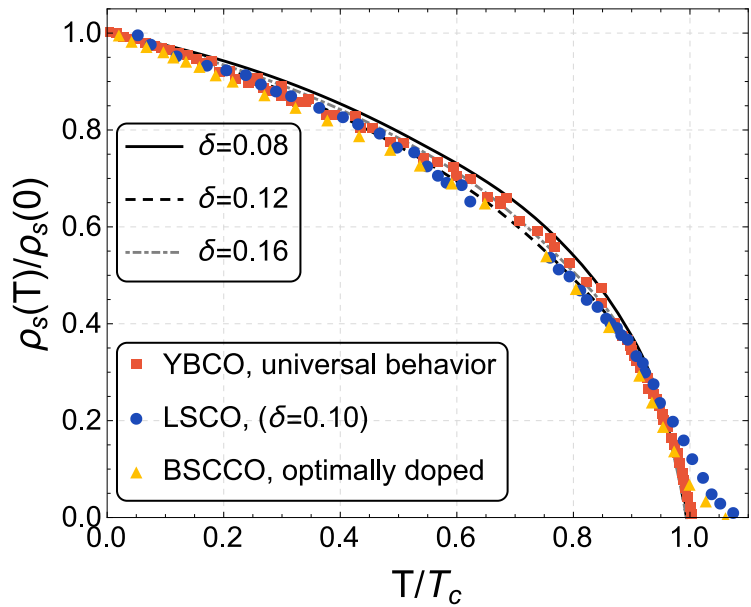}
\caption{Near-universal behaviour of superfluid density over a wide range of doping, compared with experimental data for underdoped (LSCO, YBCO) and optimally doped (BSCCO, YBCO) samples, from Refs. \cite{hardy,jacobs,panagopoulos}.}
\label{fig:2}       
\end{figure}

The holon contribution is more standard. In fact it can be directly calculated by adapting the formula for superfluid density in $d$-wave BCS superconductor as
\beq
\rho^{(s)}_h(T)=\frac{2 \epsilon_F}{\pi} \left(1-\frac{\log(2)}{2 \Delta_h} T \right) \; ,
\eeq
allowing one to calculate the final result for superfluid density by using Eq. (\ref{eq:ilrhos}). Moreover if we normalize the resulting superfluid density as
\beq
\rho^{(s)}_n (T/T_c) = \frac{\rho^{(s)}(T/T_c)}{\rho^{(s)}(T=0)} \; ,
\eeq
we observe almost universality over a broad range of doping, from moderate underdoping to optimal doping, where the spinon contribution is dominating, realizing condition 3) of universality. 
As for the resistivity the spinon contribution shows a higher degree of universality which is however partially lost when added to the holon contribution. Again it turns out that the non-universal holon contribution is needed in a more precise fitting of the experimental data, as shown in Fig. \ref{fig:2}, with the $\delta,T$-independent ratio between the two contributions as the only free parameter.

\section{Conclusions}

In this work we have analyzed the universal properties of in-plane resistivity and superfluid density for underdoped hole-doped cuprates.

Firstly a spin-charge gauge approach to superconductivity in cuprates has been introduced: the hole is decomposed as a product of a spinful, chargeless spinon and a spinless, charged holon; each particle excitation is bound to a gauge field providing a statistical flux and allowing one to modify the statistics, in particular semionic statistics are chosen for both excitations. Within an opportune mean-field approximation holons are described by a BCS-like $d$-wave Hamiltonian and pair at a temperature $T_{ph}$. The spinons, on the other hand, are described by a non-linear $\sigma$ model and pair at a lower temperature $T_{ps}$ ($\lesssim T_{ph}$), below which a finite density of incoherent spinon pairs is formed. Finally at an even lower temperature $T_c$ the phase coherence for the recomposed hole is achieved, leading to superconductivity.

One of the most inexplicable characteristic of cuprates is the interplay, observed in many different experimental measurements, between BCS-like dynamics and a non-Fermi-liquid, non-mean-field behaviour. Within the present approach this puzzle finds a natural explanation: for instance it has been demonstrated that holons are responsible for the correct symmetry of the order parameter and for the shape of the Fermi surface, while spinons are responsible for the non-mean-field critical exponent of the critical transition.

Clearly within this composite approach one expects the response functions to have a contribution arising from spinons and one arising from holons, in the most general case. How do they sum up? For quantities depending on the low-energy and low-momentum polarization bubble the Ioffe-Larkin composition rule holds under opportune assumptions, allowing one to derive a composition rule e.g. for in-plane resistivity and for superfluid density.

Within this picture we argue that if a quantity has both holonic and spinonic contributions, and if the two contributions sum according to the Ioffe-Larkin rule while the spinon contribution is dominating, then there will be a chance of finding universality with strongly non-Fermi liquid character, as a result of the quantity being essentially controlled by the spinons.

We demonstrate that within our theoretical framework the assumptions are verified for the in-plane resistivity in the normal state and for the superfluid density in the superfluid state. We discuss and analyze the universality for these two quantities, when a proper rescaling is introduced. Our theoretical findings are compared with experimental data, showing good agreement. In both cases we observe that the holon contribution is subleading: it does not determine the overall qualitative behaviour of both quantities we analyze, however it gives rise to smaller corrections needed in order to provide quantitative agreement with experiments.

Let us comment on future extensions of the present work. In the treatment of the in-plane resistivity presented in this paper we did not take into account holon and spinon pairs formation; this might be achieved by adapting the methods developed in Ref. \cite{mg} to deal with holon pairs. Extending to the superconducting state the formalism discussed in the same reference permits to deal with superfluidity at doping concentration higher than those considered here. Preliminary results on the superfluid density in the strange metal phase exhibit a larger contribution from holons and therefore a lower degree of universality.

In conclusion, the present analysis shows that the spin-charge gauge approach is able to reproduce universality observed for the in-plane resistivity and for the superfluid density. Besides fitting experimental data with good precision, the present formalism also provides a natural explanation for the universal behaviour.

\begin{acknowledgements}
For the authors it is a great pleasure to acknowledge F.~Toigo for many illuminating discussions. P.A.M.  thanks Z.~B.~Su, L.~Yu and F.~Ye for the joy of a longtime collaboration. 
\end{acknowledgements}

\end{document}